# THE SINGLE ELECTRON R-PUMP: FIRST EXPERIMENT

S. V. Lotkhov, S. A. Bogoslovsky*, A. B. Zorin and J. Niemeyer
Physikalisch-Technische Bundesanstalt
Bundesallee 100, D-38116 Braunschweig, Germany

## Abstract

We fabricated and tested the single electron R-pump, i.e. a three-junction Al circuit with on-chip Cr resistors. Due to the presence of the resistors ($R > R_K \equiv h/e^2 \approx 25.8$ k$\Omega$), the accuracy of electron transfer in the R-pump can approach the level of $\delta I/I \sim 10^{-8}$. Preliminary results of experiment with the R-pump made at PTB are reported.

## Introduction

As was demonstrated by Pothier *et al.* [1,2], the single electron pump, i.e. a chain of three junctions (with two gates related to the two islands), can transfer individual electrons, i.e. carry the current $I = ef$, when an rf signal of frequency $f$ is applied to the pump gates. In such a pump driven typically at $f \sim 10$ MHz (<< $(R_j C)^{-1}$, $R_j \sim 100$ k$\Omega$, the tunnel resistance and $C \sim 0.3$ fF, the junction capacitance) the relative accuracy of pumping, $\delta I/I$, achievable is hardly better than $10^{-2}$-$10^{-3}$. This limitation is imposed by the cotunneling effect, i.e. unwanted events of electron tunneling through several (i.e., two or even all three) junctions at the same time.

To suppress cotunneling and, in doing so, to improve the accuracy of the dc current up to the "milestone" level of $10^{-8}$, the number of junctions in the chain, $N$, should normally be $\geq 5$. [3] So far, this accuracy was approached by Keller *et al.* [4] only in a seven-junction pump. The operation of this unique device was, however, rather complex because of the large number of islands and gates (six): one has to find and maintain a working point of the pump in the 6-dimensional space of parameters. Moreover, to greater extent than its three-junction counterpart this pump suffers from fluctuations of offset charges occurring on all 6 islands.

In this paper we report the preliminary results obtained on a three-junction R-pump: a device which combines an attractive simplicity of the conventional three-junction pump and low cotunneling leakage typical of many-junction ($N > 3$) circuits. The key elements of an R-pump are the on-chip resistors of $R \sim$ several $R_K \equiv h/e^2 \approx 25.8$ k$\Omega$, which set up the dissipative environment for the junctions and, therefore, efficiently damp cotunneling in the system.

## Peculiarity in Operation

The operation principle of the R-pump (see Fig. 1) is almost the same as that of the conventional three-junction pump [1,2]. Since the condition $(RC)^{-1} >> f$ is met, charge equilibrium is established fast, just after a tunneling event and, therefore, during most of the cycle the voltage drop across the resistors is zero. The device has the same diagram of stability [1,2], namely the honeycomb pattern in the plane of variables $\{V_1, V_2\}$ at zero total voltage across the pump, $V = V_L - V_R = 0$.

The peculiarity of the R-pump dynamics consists in modified rates of electron tunneling. Compared to the conventional pump, the rate of tunneling across one junction for $R = 50-100$ k$\Omega$ is reduced several times [5]. At the same $R_j$ and $C$ this leads to a somewhat lower operation frequency $f_{max}$. On the other hand, the rate of cotunneling across two and, especially, three junctions is reduced by several orders of magnitude [6]. That is to say, for two-junction (three-junction) cotunneling the resistor roughly acts similar to $\Delta N = R/R_K$ ($\Delta N = \frac{1}{2} R/R_K$) tunnel junctions attached to a three-junction pump [7].

## Samples

The Al tunnel junctions (about 60 nm × 60 nm) and the Cr thin film resistors (7 nm thick, 80 wide and 10 µm long) were fabricated *in situ* by the shadow evaporation technique through the trilayer mask patterned by e-beam lithography and reactive-ion etching.

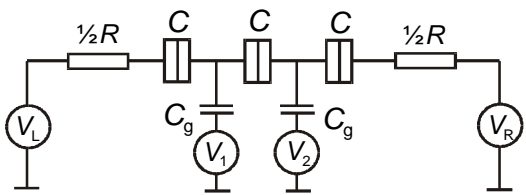

Fig. 1. The equivalent scheme of the R-pump. The voltage drop across the device ($V_L$–$V_R$) is assumed to be 0; the voltages applied to the gates $V_{1,2} = V_{10,20} + V_A \sin(2\pi f t \pm \frac{1}{2}\theta)$, where $V_{10,20}$ are the dc offset voltages at which the chain is in the unblocked state (so-called "triple-point") and the rf signal amplitude $V_A \sim 0.3\ e/C_g$. The phase delay $\theta = 90°-180°$ can compensate the unavoidable cross-coupling (not shown) so that the ac-polarization charges induced on the islands have the favorable phase shift of 90°.

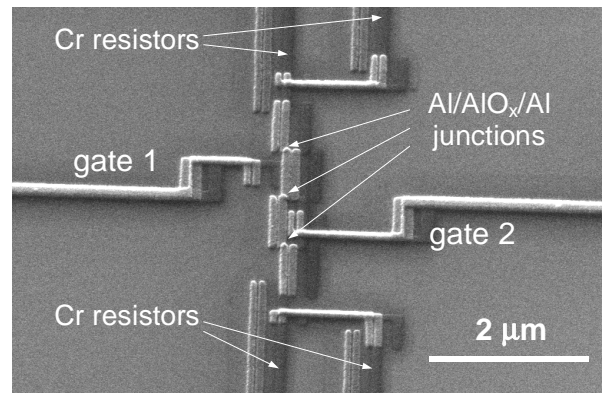

Fig. 2. SEM image of the three-junction R-pump fabricated by the three-shadow evaporation technique.

In the group of three samples tested we found the junction self-capacitances $C$ = 250–300 aF ($\gg C_g \approx$ 40 aF) and tunnel resistances $R_j$ = 100–150 k$\Omega$ (depending on sample). Four identical resistors of $R \approx$ 60 k$\Omega$ attached in pairs to the chain (see Fig. 2) were used for a 4-point measurement of the R-pump as well as for the characterization of the resistors themselves. The cross-talk capacitances (between a gate and an alien island) amounted to about 40% of the coupling capacitances $C_g$.

## Results

The measurements were carried out in a dilution fridge at the bath temperature of $T \approx$ 20 mK and in a magnetic field of 1 tesla which ensured the normal state of Al parts of the circuit. The dc and ac lines comprised "cold" pieces of the Thermocoax® cable (1 m and 60 cm long, respectively) which served as a microwave frequency filter and ensured reasonable attenuation at frequencies > 1 GHz. The total attenuation in the ac lines for the driving signal of $f$ = 10 MHz was about 15 dB. All three samples proved to have good operation characteristics. Figure 3a shows the I-V curves of one of the samples in the blocked and the open (i.e. in a triple point) state, in the presence of the rf drive and without it. The horizontal steps of current (see the zoomed up picture in Fig. 3b) caused by an rf drive of $f \leq$ 10 MHz with the phase shift of $\theta = \pm (120°–150°)$, are indicative of the electron pumping giving rise to current $I = ef$.

The sensitivity of our measuring setup was $\sqrt{S_I}$ ~ 50 fA/$\sqrt{\text{Hz}}$ and did not allow the errors of pumping to be directly measured as, for example, in Ref. 4. Therefore we evaluated these errors from the shape of the rounded corners of the step edges (see Fig. 3b). These parts of the I-V curve for $f$ = 6 MHz (as well as of the I-V curve corresponding to the blocked state) were fitted quite well by two exponential functions $A_\pm \exp(\pm eV/2k_B T^*)$ with the characteristic temperature of $T^* \approx$ 116 mK (see Fig. 3b). This relatively large value characterizes the integral action of noise of different nature (rf and black-body radiation, the 50 Hz pickup, mechanical vibrations in the system, etc.).

Taking the obtained value of $T^*$ = 116 mK into account we found the relative errors of pumping in the central part of the step to be about 5×10$^{-4}$. Through the improvement of filtering in the setup $T^*$ can be reduced down to the electron temperature of the islands, $T_e \approx$ 50–60 mK. In that case one could reach for the present samples the level of 10$^{-6}$ which can be reduced down to 10$^{-8}$ by moderate reduction of junction capacitance $C$.

As cotunneling errors are concerned, we evaluated them with the aid of Eqs.(9) and (10) of Ref. 7. For the parameters of the present samples and the frequency $f \leq$ 10 MHz, these errors, as well as those resulted from the cycle missing events, are expected to be lower than 10$^{-8}$.

## Acknowledgment

The work is supported in part by the EU (Projects CHARGE and COUNT) and the German BMBF (Grant No. 13N7168).

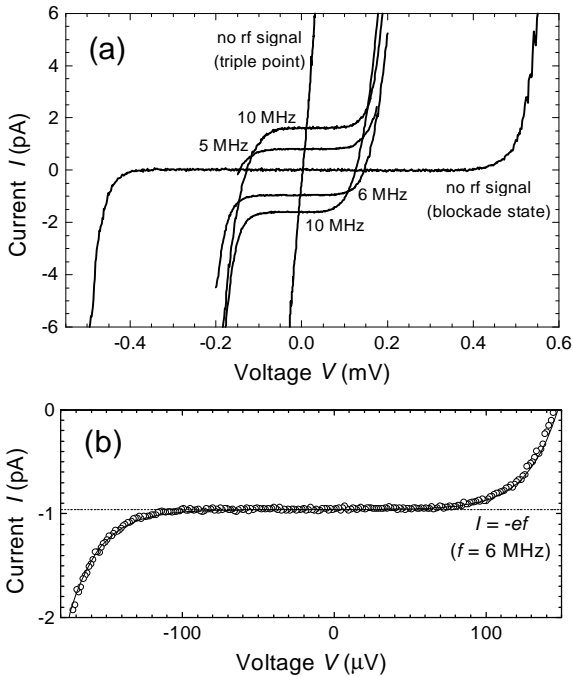

Fig. 3. I-V characteristics of R-pump at different frequencies of rf drive (a). The blowup (b) of the step at $I = -0.96$ pA (this level is shown by dotted line): symbols are data, thin solid line is the exponential fit.